\documentclass{article}
\usepackage{spconf,amsmath,graphicx}
\usepackage{algorithmic}
\usepackage{algorithm}
\usepackage{booktabs}

\title{Multi-task Voice Activated Framework using Self-supervised Learning}

\name{*Shehzeen Hussain$^1$\thanks{* Work performed as an intern at Qualcomm Technologies} \qquad Van Nguyen$^{2}$ \qquad Shuhua Zhang$^{2}$\qquad Erik Visser$^{2}$}
\address{$^{1}$ University of California, San Diego \\
      {$^{2}$}Qualcomm Technologies, Inc.,San Diego California }
\begin{document}
%
\maketitle
\begin{abstract}
Self-supervised learning methods such as wav2vec 2.0 have shown promising results in learning speech representations from unlabelled and untranscribed speech data that are useful for speech recognition. 
Since these representations are learned without any task-specific supervision,
they can also be useful for other voice activated tasks like speaker verification, keyword spotting, emotion classification etc.
In our work, we propose a general purpose framework for adapting a pre-trained wav2vec 2.0 model for different voice activated tasks. We develop downstream network architectures that operate on the contextualized speech representations of wav2vec 2.0 to adapt the representations for solving a given task. 
Finally, we extend our framework to perform multi-task learning by jointly optimizing the network parameters on multiple voice activated tasks using a shared transformer backbone. 
Both of our single and multi-task frameworks achieve state-of-the-art results in speaker verification and keyword spotting benchmarks. Our best performing models achieve 1.98\% and 3.15\%  EER on VoxCeleb1 test set when trained on VoxCeleb2 and VoxCeleb1 respectively, and 98.23\% accuracy on Google Speech Commands v1.0 keyword spotting dataset. 
\end{abstract}
\section{Introduction}
\label{sec:intro}
Currently, the best performing solutions for voice activated tasks such as keyword spotting (KWS), speaker verification (SV), emotion classification, language identification etc. require training deep neural networks on large domain-specific and labeled speech datasets~\cite{xie2019utterance,xiang2019margin,qualcommkwssota,cai2018exploring,jung2020improving}. The most commonly used models typically operate on the \textit{mel-spectrogram} representation of audio. Such models require task-specific engineering techniques like tuning the window and hop size of the short-time Fourier transform (STFT), determining the number of mel bins and various denoising transformations as input pre-processing~\cite{arias2021multi,dangol2020speech}. 
Additionally, since mel-spectrogram is a compressed audio representation, it may discard information that can potentially be useful for discriminating on voice activated tasks.

Recently proposed self-supervised learning methods such as wav2vec 2.0~\cite{wav2vec2}, can learn useful speech representations directly from unlabeled/un-transcribed speech data. 
The authors of wav2vec 2.0 found a strong correlation between the learned representations and phonemes, and demonstrated that these learned representations can achieve state-of-the-art results on speech-to-text benchmarks with much less amount of labelled data as compared to prior work. 
However, these representations could also be useful for other voice activated tasks besides speech recognition, and can potentially address the limitations of spectrogram representations when performing voice activated tasks. Solving voice activated tasks such as keyword detection and speaker verification using learned representations from self-supervised learning can reduce the amount of labeled data required for training and also result in better performing models. Moreover, since these speech representations are derived directly from the waveform, it simplifies the detection and classification pipeline by providing an end-to-end solution.

In this work, we develop a framework for multi-task voice activated systems based on contextualized speech representations learned from self-supervised models. Our framework consists of two major components \textit{1) Speech Representation Extractor (SRE):} A self-supervised model (in our case wav2vec 2.0) that provides contextualized speech representations directly from a waveform.
\textit{2) Downstream Neural Network:} To guide the training of SRE and shape the representations into solving multiple voice activated tasks such as KWS and SV. 

\begin{figure*}[htp]
    \centering
    \includegraphics[width=1.0\textwidth]{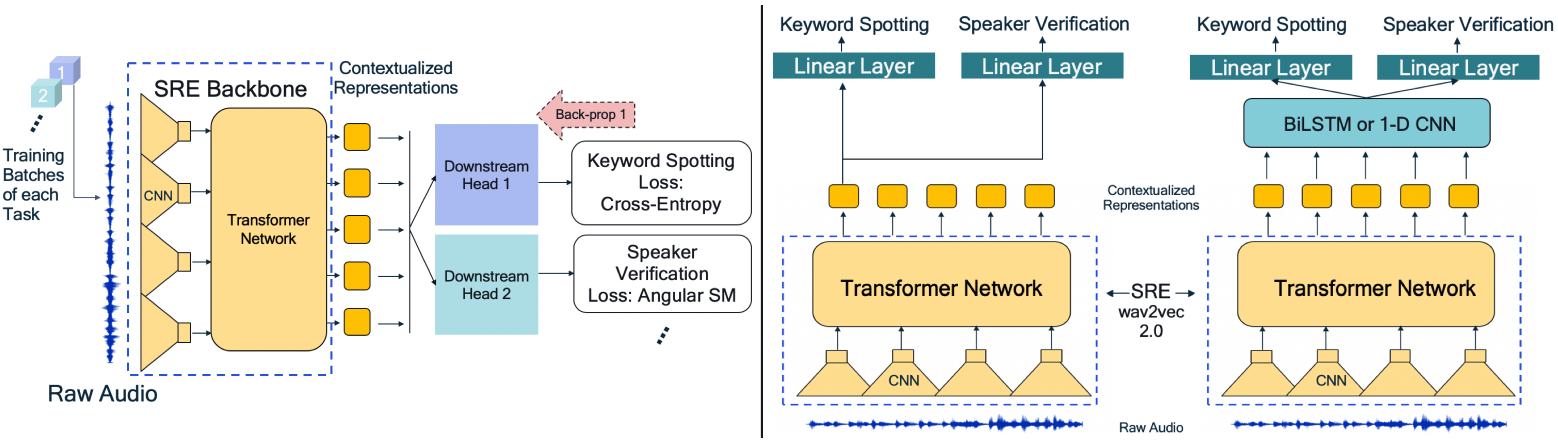}
    \vspace{-8mm}
    \caption{Overview of our proposed multi-task training framework with wav2vec 2.0 based SRE backbone and various downstream networks to solve voice activated tasks.
    }
    \label{fig:model_diagram}
\end{figure*}

In our framework, the SRE is first pre-trained in a self-supervised manner on a large unlabelled speech corpus. Then the downstream networks and SRE are jointly optimized on single or multiple tasks using supervised learning. Our results demonstrate the representations learnt using self-supervised training can boost KWS and SV performance substantially. Our main contributions are as follows:
\begin{itemize}
\itemsep0em 

    \item We achieve state-of-the-art performance on KWS and SV tasks and outperform previous works using standard optimization objectives for both tasks. 
    \item We propose a multi-task training strategy for our framework on disjoint datasets for different tasks.
    Our multi-task training framework not only reduces the memory footprint of our models, but also achieves better performance than our single-task framework.
    \item We provide a comparative analysis of different neural network architectures when finetuning Wav2Vec 2.0 on downstream tasks. Our results suggest that a computationally inexpensive linear layer can achieve competitive results as compared to temporal models like Bi-LSTM and 1D-CNN.
\end{itemize}


\section{Methodology}
\label{sec:method}

Our framework for solving voice activated tasks consists of two major components --- 1) A Speech Representation Extractor (SRE) which is trained in a self-supervised manner on raw audio data. 2) Downstream networks that operate on the speech representations extracted from the SRE to map them to a target class or an embedding for solving the voice activated task. We first describe these two components and then discuss our multi-task training algorithm that trains multiple downstream heads jointly with a common SRE backbone. 

\subsection{Speech Representation Extractor (SRE):}
Our SRE is a wav2vec 2.0 model~\cite{wav2vec2}, comprised of convolutional and transformer layers that operate directly on the raw audio waveforms. Figure~\ref{fig:model_diagram} gives a high level overview of this network. The convolutional encoder serves as a feature extractor and operates on 20 milliseconds windows of the waveform and outputs the encoded representations $\mathit{Z}$. The transformer module operates on $Z$ and produces contextualized audio representations $C$ by attending over the future and past audio frames. 
The wav2vec model is trained in a self-supervised manner by masking some of the encoded representations $Z$ before feeding them to the transformer module. The transformer module is trained with the objective of predicting the quantized version $Q$ of the encoded representations $Z$. The discrete representation $Q$ is obtained from $Z$ using a learnable quantization module. The training objective of wav2vec 2.0 uses contrastive loss where the network is encouraged to distinguish the true masked quantized representation from distractor samples of other time-steps. Additional loss terms are employed to ensure diversity in the codebooks of the quantization module and $L2$ regularization of network weights.  
For the SRE backbone of our framework, we use a pre-trained wav2vec 2.0 Base model that is trained on LibriSpeech dataset with around 1000 hours of speech. We refer the readers to~\cite{wav2vec2} for more details on wav2vec 2.0 pre-training.

\vspace{-3mm}
\subsection{Downstream Networks:}
Downstream networks are needed to map the output of the SRE to task specific outputs. Fine-tuning of the wav2vec 2.0 model was originally explored for speech recognition in which the output of the wav2vec model at each time-step is mapped to the scores over the characters/phonemes in the vocabulary using a linear layer and optimized using CTC~\cite{graves2006connectionist} loss. 
However, for tasks like KWS, SV, etc. the entire utterance has to be mapped to a single class (for classification tasks) or an embedding vector (for tasks like speaker verification). To achieve this mapping, we need a learnable module that can operate on the outputs of the SRE at different time-steps. 
In our multi-task learning framework we have one downstream network for each task.
We explore the following downstream networks to achieve this mapping:

\noindent \textbf{1) Linear Layer:} A Linear layer is our most light-weight downstream network in which we map the output of the SRE at the first-timestep to the class scores or an embedding vector using a linear layer. Since the SRE is a transformer-based neural network, temporal dependencies on the future audio frames can be learned by the trainable attention modules during fine-tuning. 

\noindent \textbf{2) Bidirectional LSTM (Bi-LSTM):} In this setup, we have a bi-directional LSTM that operates over the vector outputs at each time-step of the SRE. Then the output of the first and last-timestep of the Bi-LSTM are combined and mapped to the task-specific output using a linear layer. We use a single-layer BiLSTM with 256 hidden units.
    
\noindent \textbf{3) 1D-CNN:} 
Our 1D-CNN comprises of strided convolutional blocks with filters of size 25. Each convolutional block consists of a convolutional layer with 128 filters followed by ReLU activation and a batch normalization layer.
We use two CNN architectures for one second and two seconds audio input respectively. For one second audio, we use a single convolutional block with stride 4 to reduce the SRE output timesteps to 16. We flatten this final output before feeding it to a linear layer to map it to our task-specific output. 
For two seconds utterances, we have an additional convolutional block with stride 2 as the first block, followed by the same architecture as that used for one second audio. 
\subsection{Multi-task Training:}


In our multi-task learning framework, we train a shared SRE backbone with multiple downstream heads, with each downstream head corresponding to a different voice-activated task. Multi-task learning has two main advantages --- 1) A shared SRE backbone across multiple tasks reduces the memory footprint and makes our setup suitable for deployment in memory-constrained settings. 
2) Features useful for discriminating on one voice activated task can also potentially be useful for alternate tasks.
Performance gains due to multi-task learning have been demonstrated in the image domain~\cite{xiang2017joint,9427073,deng2020retinaface,lin2017focal}.

\begin{algorithm}
   \caption{Multi-Task Training Algorithm}
   \label{alg:multitask}

\begin{algorithmic}
    \STATE $\textit{model}$ $\gets$ Initialize SRE + Downstream Model 
        \STATE $\textit{optimizer}$ $\gets$ $\textit{AdamOptimizer(model)}$
    \FOR{\textit{task} in \textit{AllTasks}}
        \STATE \textit{Initialize} $\mathit{DataLoader}_{\mathit{task}}$
        \STATE \textit{Define} $\mathit{Criterion}_{\mathit{task}}$
    \ENDFOR
    
    \FOR{ $\mathit{iter}$ in $[0, MaxIterations]$}
    
         \FOR{\textit{task} in \textit{AllTasks}}
            \STATE $x$, $y$ $\gets$ \textit{next}($\mathit{DataLoader}_{\mathit{task}}$)
            \STATE $y_\textit{pred}$ $\gets$ $\textit{model(x)}$
            \STATE $\textit{loss} \gets \textit{Criterion}_{\textit{task}}(y_\textit{pred}, y)$
            \STATE $\textit{BackProp}(\textit{loss})$
            \STATE \textit{\textit{\textit{optimizer}.\textit{update}()}}
        \ENDFOR
    \ENDFOR
   
\end{algorithmic}
\end{algorithm}

Multi-task training setup is straightforward if the target classes/outputs for all the tasks are present for any given utterance in our dataset.
However, most large-scale public datasets for voice activated tasks are disjoint. For example SV datasets like VoxCeleb~\cite{nagrani2017voxceleb} do not contain keyword information and vice-versa. Therefore, we use our training mechanism described in Algorithm~\ref{alg:multitask} for learning our multi-task framework. We initialize the SRE model parameters from the pretrained wav2vec 2.0 checkpoint trained in a self-supervised manner and randomly initialize our downstream networks.
We then initialize mini-batch dataloaders and loss criteria for each task. We use Cross-Entropy (CE) loss for KWS and angular softmax loss~\cite{Liu_2017_CVPR} with cosine similarity on the embeddings for SV task. 
In our training loop, we go through the mini-batches for each task in a round-robin manner. We keep the SRE backbone frozen for the first one thousand mini-batch iterations and then unfreeze it to jointly update the SRE parameters with the downstream networks. We use Adam optimizer with a learning rate of $10^{-4}$ for the downstream network parameters and $10^{-5}$ for the SRE backbone. We use a lower learning rate for SRE to prevent overfitting during finetuning. 

\section{Experiments}
\subsection{Datasets}

\noindent \textbf{KWS:} For keyword spotting we choose the widely used Google Speech Command (GSC) Datasets V1 and V2~\cite{warden2018speech}. 
There are total of 30 keywords and we use ten classes of "Yes", "No", "Up", "Down", "Left", "Right", "On", "Off", "Stop",
and "Go" with two additional classes "Unknown Word (the remaining 20 words)" and "Silence (no speech detected)" following the settings of~\cite{qualcommkwssota,seo2021wav2kws}.
This results in GSC1-12 and GSC2-12 with shared 12 commands from datasets V1 and V2. Both of these datasets contain 1 second utterances.
We add the same data augmentation and noise mixing for the training samples as~\cite{seo2021wav2kws}, use the same data-split as used in in~\cite{qualcommkwssota,warden2018speech,seo2021wav2kws} and report the Top-1 Test accuracy on the standard test sets.

\noindent \textbf{SV:} For the speaker verification task we use VoxCeleb-1 and VoxCeleb-2 datasets, which are large-scale text-independent speaker recognition datasets containing 1,211 and 5,994 speakers respectively in training set~\cite{voxcelebwildjournal,nagrani2017voxceleb}. We use 2 second utterance slices for training our networks. Neither voice activity detection (VAD) nor data augmentation is used for SV training. 
Evaluations for both VoxCeleb-1 and VoxCeleb-2 are conducted on the full 37611 trial pairs of 40 unseen speakers from the original test set of VoxCeleb-1~\cite{voxcelebwildjournal} available on their website~\footnote{https://www.robots.ox.ac.uk/~vgg/data/voxceleb/vox1.html}.The speaker identities in testing set are disjoint from the ones in training set. We evaluate the Equal Error Rate (EER)~\footnote{Error at the threshold where Fase Positve Rate=False Negative Rate} metric on these trial pairs using 256 dimensional speaker embeddings.

Table~\ref{tab:datasets} lists the characteristics of individual datasets for both KWS and SV used in our experiments. 

\begin{table}[htp]
\centering
\resizebox{0.8\columnwidth}{!}{%
\begin{tabular}{l|rrr}
\toprule
Dataset & \# Utterances & \# Hours & \# K  \\ 
\midrule
GSC1 - (12) Train & 22236 & 6.2 & 12 \\
GSC2 - (12) Train & 30769 & 8.6 & 12  \\
GSC1 - (12) Test & 3081 & 0.9 & 12 \\
GSC2 - (12) Test & 4890 & 1.4 & 12 \\
\midrule
VoxCeleb1 Train & 143642 & 329.06 & 1211 \\
VoxCeleb2 Train  & 1021161 & 2207 & 5994\\
VoxCeleb1 Test  & 4874 & 11.2 & 40 \\
 \bottomrule
\end{tabular}%
}

\caption{Statistics of the datasets used in our experiments. We use the GSC datasets for KWS and VoxCeleb datasets of SV. \#K refers to the number of speakers for SV and number of keywords for the GSC dataset.
}
\label{tab:datasets}
\end{table}




\subsection{Results}
\noindent \textbf{Single-Task:}
In the single-task setup we train our models to solve individual voice activated tasks, that is, separate models for KWS and SV. 
For KWS, we trained our SRE and linear downstream network heads on both versions of Google Speech Commands dataset using a single-task framework.
Table~\ref{tab:multitask} lists the Top-1 Accuracy of experiments with the 12 shared commands in datasets V1 (GSC1-12) and V2 (GSC2-12). We compare the performance of our models against prior state-of-the-art results in Table~\ref{tab:multitask}.
While our multi-task framework achieves higher performance on KWS when compared to our single task framework, it is worth noting that even our single task framework achieves better performance than previous SOTA models. 


For SV, we train two separate models on VoxCeleb-1 and Voxceleb-2 training sets and evaluate them on the unseen Vox-1 test set. We report the Equal Error Rate (EER) for this evaluation in Table~\ref{tab:multitask}. Similar to KWS, our single task framework for SV also achieves better performance than previous SOTA. Even when training with only VoxCeleb-1 dataset we outperform previous baselines~\cite{exploringSVLID,2018attentive} that have trained with same dataset (3.35\% EER compared to 3.61\%). 

\noindent \textbf{Multi-Task:}
We present the results of our multi-task framework on voice activated tasks KWS and SV in Table~\ref{tab:multitask}. The results indicate that training tasks together can yield improvement over training them separately, as demonstrated by the higher Top-1 Accuracy for KWS and lower EER for SV achieved by multi-task setup compared to the single task setup in Table~\ref{tab:multitask}. 
Our proposed multi-task framework enables shared SRE backbone to solve multiple tasks thereby reducing memory footprint when solving multiple tasks without hindering individual task accuracy. 
With our proposed framework, we achieve state-of-the-art performance on both KWS and SV voice activated tasks on all benchmark datasets explored in our work. 

We perform two additional experiments: 1) Using an untrained randomly initialized SRE that is trained from scratch with the downstream network 2) using a pre-trained (no finetuning) Frozen SRE. The results for these experiments reported in Table~\ref{tab:multitask} ascertain the importance of both pre-training and finetuning the SRE backbone. 
 
\setlength\tabcolsep{2 pt}
\begin{table}[htp]
\centering
\resizebox{\columnwidth}{!}{%
\begin{tabular}{@{}l|l|cc@{}}
\toprule
 & Train Dataset & SV-EER(\%) & KWS-Acc(\%) \\ \midrule
CNN + Embedding~\cite{voxcelebwildjournal} & VoxCeleb1 & 7.80 & - \\
Attentive statistics~\cite{2018attentive} & VoxCeleb1 & 3.85 & - \\
Wav2Vec-SV~\cite{exploringSVLID} & VoxCeleb1 & 3.61 & - \\
ResNet+AM Softmax~\cite{xie2019utterance} & VoxCeleb2 & 3.23 & - \\
CNN+GhostVLAD~\cite{voxcelebwildjournal} & VoxCeleb2 & 2.87 & - \\
DNN+AAM Softmax~\cite{xiang2019margin} &VoxCeleb2 & 2.69 & - \\
Wav2KWS~\cite{seo2021wav2kws} & GSC1 (12) & - & 97.9 \\
BC-ResNet-8-KWS~\cite{qualcommkwssota} & GSC1 (12) & - & 98.0 \\ \midrule

SRE + Linear single & GSC1 (12) & - & 98.10 \\
SRE + Linear single & GSC2 (12) & - & \textbf{99.28} \\
SRE + Linear single & VoxCeleb1 & 3.35 & - \\
SRE + Linear single & VoxCeleb2 & 2.08 & - \\ 
\midrule
SRE + Linear multi & VoxCeleb1 + GSC1 (12) & \textbf{3.15} & 98.19 \\
SRE + BiLSTM multi & VoxCeleb1 + GSC1 (12)& 3.53 & 98.19 \\
SRE + CNN multi & VoxCeleb1 + GSC1 (12)& 3.28 & 98.11 \\
SRE + Linear multi & VoxCeleb2 + GSC1 (12)& \textbf{1.98} & 98.23 \\
SRE + BiLSTM multi & VoxCeleb2 + GSC1 (12)& 2.31 &  \textbf{98.25}\\
SRE + CNN multi & VoxCeleb2 + GSC1 (12)& 2.52 & 98.21\\ 
\midrule
SRE (Random) + Linear  & VoxCeleb1 + GSC1 (12)& 13.2 & 96.68 \\
SRE (Frozen) + Linear  & VoxCeleb1 + GSC1 (12)& 20.5 & 90.33\\ 
\bottomrule
\end{tabular}%
}
\caption{Results of our multi-task learning experiments using different downstream heads and dataset combinations, and performance of prior state of the art methods on the datasets.
}
\label{tab:multitask}
\end{table}

\section{Relation to Prior Works}
Concurrent works have shown that self-supervised learning models like wav2vec 2.0 can be fine-tuned to perform various voice activated tasks such as KWS~\cite{seo2021wav2kws}, SV~\cite{exploringSVLID}, language identification~\cite{exploringSVLID}, emotion recognition~\cite{pepino2021emotion}.
We demonstrate that our models outperform these previous approaches on two benchmark tasks: KWS and SV when using both single-task and multi-task frameworks. 
Earlier work~\cite{jung2020improving} performing multi-task learning for KWS and SV using mel-spectrograms, reports that multi-task learning can improve performance over single-task setup, and we find our results to corroborate this finding when using an end-to-end solution with contextualized speech representations from wav2vec. These results are in contrast to another attempt~\cite{exploringSVLID} to jointly learn speaker verification and language identification, which claims that multi-task learning can negatively impact single-task performance. We conjecture that this difference in gains in~\cite{exploringSVLID} from multi-task learning may be due to the chosen tasks that were solved jointly in their multi-task framework and the difference in their training algorithm.

\vspace{-3mm}
\section{Conclusion}
In this work, we propose a general purpose training framework for voice activated multi-tasks using self-supervised models. We outperform both prior methods using neural networks on mel-spectrogram representations and prior attempts at using waveform representation of speech. Our multi-task training strategy, not only reduces the memory footprint by solving multiple tasks using a shared backbone, but also improves the performance over single-task models.

\bibliographystyle{IEEEbib}
\bibliography{bib}

\end{document}